\documentclass[aps,prl,superscriptaddress,showpacs,twocolumn]{revtex4}
\usepackage{amsfonts}
\usepackage{amsmath}
\usepackage{amssymb}
\usepackage{graphicx}
\begin{document}
\title{Weak Plaquette Valence Bond Order in
the $S=1/2$ Honeycomb $J_1-J_2$ Heisenberg Model}
\author{Zhenyue Zhu}
\affiliation{Department of Physics and Astronomy, University of California,
Irvine, California 92697, USA}
\author{David A. Huse}
\affiliation{Department of Physics, Princeton University, Princeton, New Jersey 08544, USA}
\author{Steven R. White}
\affiliation{Department of Physics and Astronomy, University of California,
Irvine, California 92697, USA}

\begin{abstract}
Using the density matrix renormalization group, we investigate the
$S=1/2$ Heisenberg model on the honeycomb lattice with first ($J_1$)
and second ($J_2$) neighbor interactions. We are able to study long
open cylinders with widths up to 12 lattice spacings.  For $J_2/J_1$
near $0.3$, we find an apparently paramagnetic phase, bordered by an
antiferromagnetic phase for $J_2\lesssim 0.26$ and by a valence bond
crystal for $J_2\gtrsim 0.36$. The longest correlation length that
we find in this intermediate phase is for plaquette valence bond
(PVB) order.  This correlation length grows strongly with cylinder
circumference, indicating either quantum criticality or weak PVB
order.
\end{abstract}

\date{\today}
 \maketitle
\medskip

Progress in finding realistic model quantum Hamiltonians with
spin-liquid (SL) ground states has accelerated dramatically in the
last two years, almost 40 years since Anderson first proposed a
resonating valence bond state as a possible ground state of the
triangular Heisenberg model \cite{ARVB}. One key recent advance was
the discovery using the density matrix renormalization group (DMRG)
of a gapped SL ground state in the spin-1/2 kagome Heisenberg
antiferromagnet \cite{kagome, kagome2}. Spin-liquid phases have been
suggested for various other models, such as the half-filled
honeycomb Fermi-Hubbard model \cite{hubhon} and the square lattice
spin-1/2 Heisenberg antiferromagnet with second-neighbor ($J_2$)
interactions \cite{squa1, squa2}. However, some skepticism has been
expressed about the evidence for spin liquids in the latter two
models \cite{hubhon2, jq}.

The main defining feature of a quantum spin liquid is the absence of
any spontaneously broken symmetry, particularly either magnetic or
valence bond order. Frustration, which discourages order, is a key
ingredient of models potentially containing spin-liquid phases. Spin
liquids arise in several analytic treatments and exactly solvable,
simplified, but less realistic models \cite{SL}.  A key feature
distinguishing types of spin liquids is the presence or absence of a
gap to all excitations. The kagome Heisenberg spin liquid is found
to be gapped.  To satisfy the Lieb-Schultz-Mattis theorem, gapped
spin liquids for models with a net half-integer spin per unit cell
must have ``hidden'' topological degeneracies in the thermodynamic
limit, which depend on the topology of the system. The simplest
possibility is a $Z_2$ spin liquid.  Since local measurements cannot
identify $Z_2$ or other topological orders, it is challenging to
identify its presence in a numerical study.  The degeneracies
characteristic of a 2D gapped $Z_2$ spin liquid have not been
accessible for the system sizes studied to date. Odd-width cylinders
spontaneously dimerize in a pattern that is characteristic of a
quasi-one-dimensional system \cite{kagome, squa1}. Another key
feature of a $Z_2$ spin liquid is the presence of a $-\ln 2$
constant term correction to the linear growth of the entanglement
entropy with a subsystem perimeter. This term has now been measured
in the nearest-neighbor kagome system \cite{kagome2} and also in the
kagome system with next-nearest-neighbor interaction $J_2$
\cite{tee}, where for $J_2=0.1$ the gaps are large and the
entanglement entropy correction term can be measured particularly
precisely. Thus, there is now solid evidence that the ground state
of the kagome spin-1/2 antiferromagnet is a gapped $Z_2$ spin
liquid.

\begin{figure*}
\includegraphics*[width=17cm, angle=0]{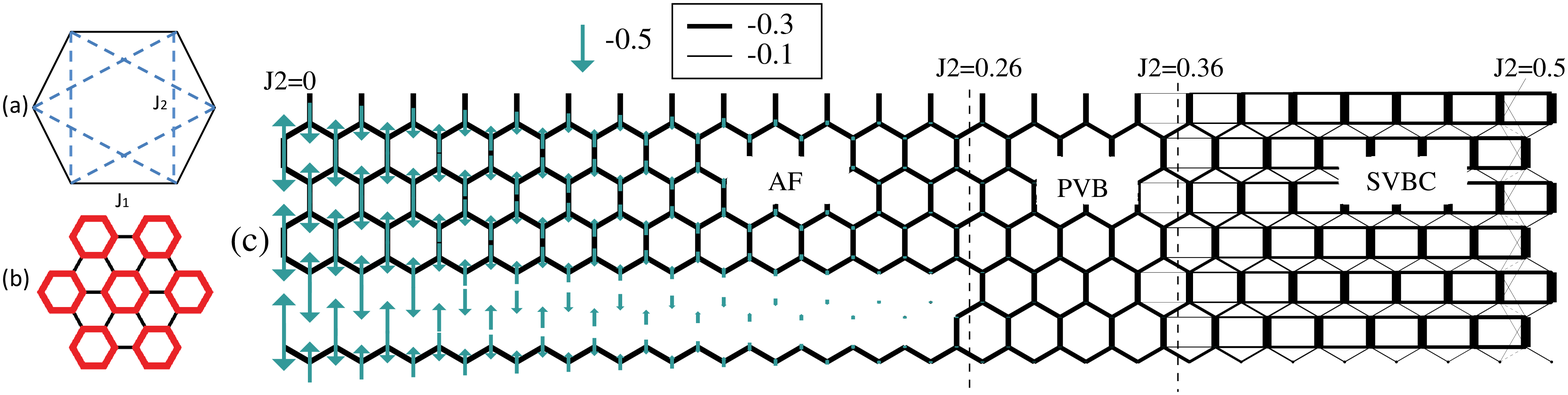}
\caption{(a) One hexagon of the honeycomb lattice showing $J_1$ and
$J_2$ interactions. The six sides of the hexagon are $J_1$ bonds,
and the six $J_2$ bonds are shown as dashed (blue) lines. (b)
Illustration of the pattern of PVB order. The spin-spin correlations
differ between the bonds shown as thin (black) lines versus thick
(red) lines. (c) Three phases on the YC6-0 cylinder with 300 sites
and a gradient of $J_2$. The widths of lines are proportional to
$|\langle S_i\cdot S_j\rangle|$. We also show the second-neighbor
spin correlation but only when its value is negative. The arrows
represents the values of $\langle S_i^z\rangle$ at each site. The
scales for correlations and magnetizations are indicated. We remove
several bonds in one zigzag row of the cylinder, so that the AF spin
pattern is more clearly seen. The two vertical lines at $J_2=0.26$
and 0.36 indicate estimates of the phase boundaries.} \label{phase}
\end{figure*}

In this letter, we examine the spin-1/2 Heisenberg antiferromagnet on
the honeycomb lattice [see Fig. \ref{phase}(a)] with Hamiltonian
\begin{equation}
H=J_1\sum_{\langle i,j\rangle}S_i\cdot S_j+J_2\sum_{\langle\langle
i,j\rangle\rangle}S_i\cdot S_j~.
\end{equation}
where the sum over $\langle i,j\rangle$ runs over nearest-neighbor
pairs of sites and the sum $\langle\langle i,j\rangle\rangle$ runs
over next-nearest neighbors. We take $J_1=1$ [antiferromagnetic
(AF)] and consider only $J_2>0$. Our work follows other studies of
this and similar Heisenberg models \cite{ed,PVB,ccm,sse,frg,RVBSL,
spiral,sw,eps,vmc}, motivated by the Hubbard model results
\cite{foot}. Most of these studies report a nonmagnetic phase near
$J_2/J_1\sim0.2-0.4$, and we agree, but the results are in general
disagreement on the range and nature of this phase. A variational
Monte Carlo study indicated a spin liquid in the range from 0.08 to
0.3 \cite{vmc}. A combination of exact diagonalization and valence
bond treatment reported a plaquette valence bond [PVB, see Fig.
\ref{phase}(b)] crystal in the range 0.2-0.4 \cite{ed}. Exact
diagonalization on small lattices \cite{PVB} and the coupled-cluster
method both suggest PVB order \cite{ccm}. Earlier work reported that
dimer correlations aren't strong enough for PVB order \cite{RVBSL}.
Functional renormalization group work claims that this phase has
weak dimer and plaquette responses \cite{frg}. Variational entangled
plaquette states suggest that none of the order parameters remain
nonzero \cite{eps}. Other theoretical work has focused on the
possibility of a $Z_2$ SL on the honeycomb lattice and on phase
transitions between Neel and staggered valence bond crystal (SVBC)
phases \cite{sb,sf,svbc}.

Here, we report that the ground state displays an apparently
paramagnetic phase for $0.26\lesssim J_2\lesssim 0.36$. For
$J_2\lesssim 0.26$, we find the usual two-sublattice AF phase. For
$J_2\gtrsim 0.36$, we find a SVBC. We have studied in some detail
the system with $J_2=0.3$, deep within the intermediate phase that
is neither AF nor SVBC.  We examine various correlation functions of
this ground state on various cylinders.  The longest correlation
length that we find is for PVB order.  The strong growth of the PVB
correlation length with cylinder circumference indicates that the
system is either near a quantum critical point or may have weak
long-range PVB order.

We use cylindrical (C) boundary conditions with open ends for our
DMRG \cite{dmrg1,dmrg2} calculations. We label the cylinders either
XCM-N or YCM-N.  The labels X or Y indicated whether a
first-neighbor bond is oriented horizontally (X) or vertically (Y).
For XC cylinders, M is the number of sites along a zigzag vertical
column and N means that the periodic boundary conditions are
connected with a shift of N zigzag columns to the left or right. For
YC cylinders, M is the number of zigzag horizontal rows and N means
the connection has a shift by N sites along a zigzag row. For
example, in the XC8-0 cylinder, one set of edges of each hexagon
lies along the X direction, and there are eight sites along the
zigzag columns, which are connected periodically. So, the
circumference is $C=4\sqrt{3}$ lattice spacings. For the YC4-0
cylinder, one set of edges of each hexagon lies along the Y
direction and the cylinder is connected periodically along the Y
direction with circumference $6$ lattice spacings. For the XC9-1
cylinder, the connection has a horizontal shift of one zigzag
column, producing a circumference of $C=3\sqrt7$.

In Fig. \ref{phase}(c), we present the ground state of a single
system which gives an overview of the entire phase diagram. For this
long YC6-0 cylinder, $J_2$ is uniform along the vertical direction,
but varies linearly with the horizontal position from $J_2=0$ at the
left edge to $J_2=0.5$ at the right edge. To make the AF order
visible, we apply a staggered field at the left end of the cylinder.
As $J_2$ increases along the cylinder, the AF order decreases,
becoming negligible in the intermediate phase. We will discuss this
intermediate state in detail below. The SVBC phase appears clearly
for $J_2\gtrsim 0.36$. This SVBC phase has strong first-neighbor
correlations along the vertical direction with strong horizontal
second-neighbor correlations connecting them to form ``ladders''.
Below we will determine the phase boundaries of the AF and SVBC
phases more accurately.

\begin{figure}
\includegraphics[width=6cm, angle=0]{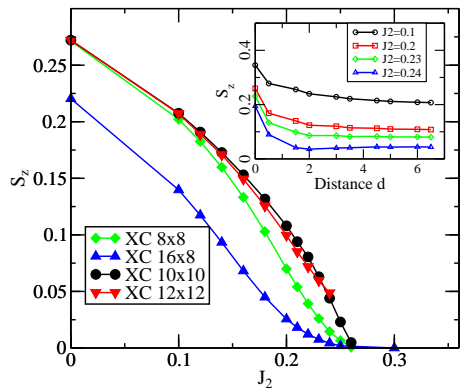}
\caption{The staggered magnetization at the center of the cylinder
versus $J_2$ for various XC cylinders. The inset shows how the local
magnetization decays from the edge of a long XC10 cylinder for
various values of $J_2$.} \label{mag}
\end{figure}

First, we estimate the boundary of the AF phase. One technique to
determine magnetic order parameters using DMRG is to put strong
ordering fields on the edges of an open cylinder, and to adjust the
aspect ratio $L_y/L_x$ to to minimize the finite-size effects
\cite{mag}. For both square and triangular spin-1/2 Heisenberg
antiferromagnets, an aspect ratio near $1.7\sim1.9$ is found to
minimize the finite-size effects.  For XCM-0 cylinders with M
columns, the aspect ratio is $\sqrt{3}$, which we use.  For $J_2=0$,
we determine that $\langle S_z\rangle\cong 0.2720$, which is close
to the value determined using Monte Carlo calculations in the
thermodynamic limit $\langle S_z\rangle=0.2677(6)$ \cite{maghon}.
With $J_2$ increasing, we find that the magnetization reduces to
near zero for $J_2 \cong 0.26$ in Fig. \ref{mag}. The various
cluster sizes all point to the phase transition near 0.26. This
phase transition point is larger than the classical limit value of
$J_2=\frac{1}{6}$. Reference \cite{vmc} claims that the Neel order
disappears at 0.08; however, the value we find here is more
consistent with other studies which give $J_2\cong 0.2$
\cite{ed,sse}.

\begin{figure}
\includegraphics*[width=6cm, angle=0]{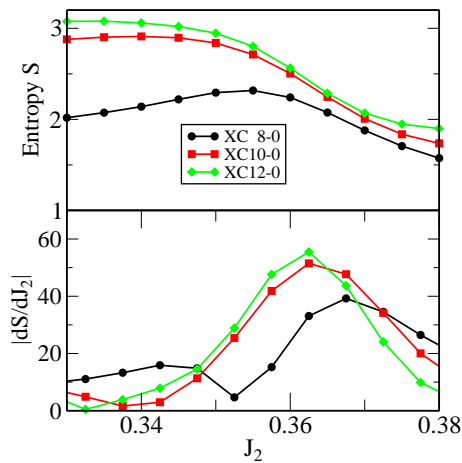}
\caption{The entanglement entropy and its derivative versus $J_2$
for different XC cylinders near the transition into the SVBC
phase.}\label{entro}
\end{figure}

To estimate the boundary of the SVBC phase we study several XC
cylinders with $J_2$ varying from 0.28 to 0.40 using the method in
Fig. \ref{phase}. These results show the SVBC phase for $J_2\gtrsim
0.36$. We also use the entanglement entropy and its first derivative
with respect to $J_2$ to estimate the phase transition point. We
make vertical cuts between zigzag columns, the dividing line between
the two parts of the system bisecting a column of horizontal bonds,
and measure the entanglement entropy. As seen in Fig. \ref{entro},
the entropy drops in going from the intermediate phase to the SVBC.
The derivative of the entropy shows a peak around 0.37 for the XC8-0
cylinder, and around 0.36 for the wider XC10-0 and XC12-0 cylinders.
In addition, the height of this peak increases with the system
width, as expected for a peak indicating a phase transition
\cite{ent,ent1,ent2}.

On XC cylinders, we find that the SVBC state has strong first- and
second-neighbor correlations along diagonal directions, forming
diagonally oriented ladders. Thus, there are two degenerate diagonal
SVBC states on an XC cylinder, whereas for YC cylinders there is
only the one vertical SVBC pattern (Fig. 2).  For an infinite
two-dimensional system, all three of these SVBC ground states would
be degenerate by rotational symmetry. In the classical limit, for
large $J_2$ values, the ground state is a spin spiral state.
However, quantum fluctuations are strong enough to melt the spiral
order and form the SVBC, in agreement with Ref. \cite{spiral}.

In the rest of this letter, we focus specifically on $J_2=0.3$
inside the intermediate phase \cite{foot1}. To measure magnetic
correlations, we apply ``pinning'' magnetic fields at one end of the
cylinder and measure the resulting magnetization pattern. Unlike in
the AF phase, the induced magnetization decays exponentially from
the end of the cylinder with a decay length of 2 to 3 lattice
spacings for various cylinders.  We further check the response to a
local magnetic field applied to a spin at the cylinder center. This
local magnetic field response is quite short ranged. It only
influences its nearby surrounding sites, as opposed to generating a
large region of staggered magnetization in the AF phase.

\begin{figure}
\includegraphics*[width=8.5cm, angle=0]{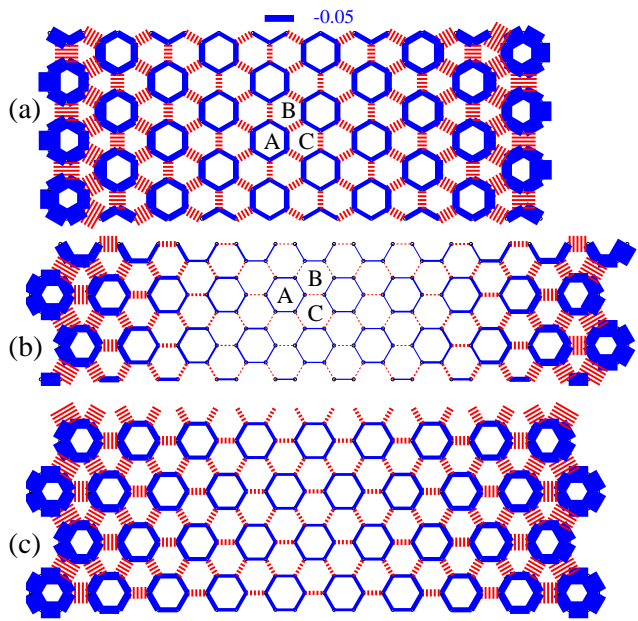}
\caption{Weak PVB order at $J_2=0.3$ on (a) YC7-3, (b) XC9-1 and (c)
XC12-0 cylinders. The widths of the lines are proportional to
$|\langle S_i\cdot S_j\rangle+0.32|$ for all the plots, with solid
blue(dashed red) lines for negative(positive) values. The bond
strength scale is indicated at the top. The three different
sublattices of plaquettes are labeled as A, B, C. In these figures,
we keep $m=6000$ states in our DMRG calculation. The truncation
error for XC9-1 is smaller than $10^{-7}$, while it is near
$10^{-6}$ for the wider YC7-3 and XC12-0 cylinders.}\label{pvb}
\end{figure}

References \cite{ed,PVB} suggest that the intermediate phase is a
PVB phase with long-range dimer-dimer correlations. To investigate
PVB ordering we study cylinders with periodic boundary conditions
that are compatible with PVB order, including YC4-0, YC6-0, YC7-3,
YC8-0, XC6-0, XC9-1 and XC12-0 \cite{foot1}.  We pin the PVB pattern
at the cylinder ends by the choice of which spins are kept and how
long the cylinder is (see Fig. \ref{pvb}).

\begin{figure}
\includegraphics*[width=7cm, angle=0]{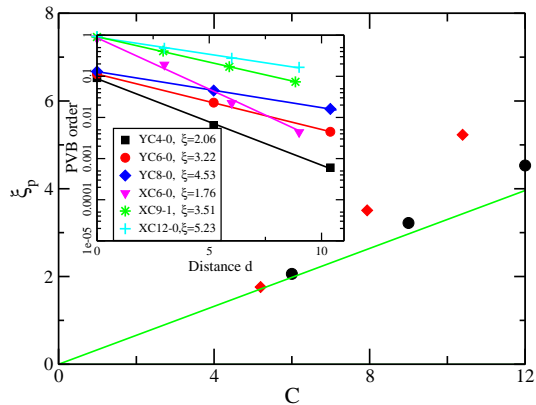}
\caption{PVB order correlation length $\xi$ for different cylinder
circumferences $C$ at $J_2=0.3$.  The black circles indicate YC
cylinders, while the red diamonds indicate XC cylinders. The
straight green line illustrates the expected linear behavior at the
quantum critical point of the two-dimensional system. The inset
shows the local PVB order parameter versus distance from the end of
the cylinder in lattice spacings. The estimated PVB correlation
length $\xi_p$ is indicated for each cylinder.}\label{pdecay}
\end{figure}

We define a local PVB order parameter at each site using the
nearest-neighbor spin correlations on the three adjacent plaquettes.
The plaquettes form three sublattices in the PVB phase, as labeled
in Figs. \ref{pvb}(a) and \ref{pvb}(b), and one plaquette from each
sublattice is adjacent to each site.  Let $E_A=-\sum_{A}\langle S_i
\cdot S_j\rangle$ be the sum over the six bonds around plaquette A.
Then, we define the local PVB order parameter as
\begin{equation}
P=E_A+E_B \exp{(\frac{2\pi}{3}i)}+E_C \exp{(\frac{4\pi}{3}i)}~.
\end{equation}
Near the ends of the cylinders, this order parameter is nonzero, and
in general it is a complex number.  We extrapolate this local order
parameter versus the truncation error to estimate its values in the
limit of large bond dimension $m$.  For long cylinders, its
magnitude decays exponentially with distance from the end, with a
correlation length $\xi_p$ that depends on the cylinder, as shown in
the inset of Fig. \ref{pdecay}. These PVB correlations can be
slightly incommensurate, particularly for the narrower YC cylinders.
For the XC9-1 cylinder with a shifted connection, we measure the
distance from the end along the direction perpendicular to the
wrapping vector to obtain the shortest PVB correlation length
$\xi_p$.

The PVB correlation length $\xi_p$ versus cylinder circumference $C$
is shown in Fig. \ref{pdecay}.  This figure includes cylinders with
all orientations, and we see that for our larger circumferences, the
XC cylinders appear to have a longer PVB correlation length than the
YC cylinders.  If the 2D system is at a quantum critical point, the
correlation length is expected to be proportional to the
circumference by standard finite-size scaling.  It appears that the
correlation length actually increases faster than the circumference,
suggesting that this system may have weak PVB long-range order in
the 2D limit of infinite circumference.

In conclusion, we have studied the $S=1/2$ honeycomb $J_1-J_2$
Heisenberg model on various cylinders extensively using DMRG. We
find that the ground state displays a paramagnetic phase for
$0.26\lesssim J_2\lesssim 0.36$. By studying PVB order on various
cylinders, we find that the PVB correlation length grows at least
linearly with the cylinder circumference. This suggests that in this
phase the system is either quantum critical or has weak long-range
PVB order. These results are compatible with an early theoretical
study that a direct phase transition between an AF and a PVB state
is possible on the honeycomb lattice \cite{qpt1}.

\begin{acknowledgments}
We thank Hong-cheng Jiang, Miles Stoudenmire, Simeng Yan, Shoushu
Gong, Olexei Motrunich, Donna Sheng and Tarun Grover for helpful
discussion. We would also like to thank A. M. L\"{a}uchli for
providing their GS energy on a small torus. This work is supported
by NSF Grants No. DMR-0907500, No. DMR-1161348 and No. DMR-0819860.
\end{acknowledgments}

\textit{Note added.}---Recently we learned that Ganesh \textit{et
al.} \cite{gane} have studied the same model with DMRG. They
reported that the ground state has three phases, including Neel
phase, f-wave PVB state phase and dimer state phase, with critical
points at $J_2/J_1=0.22$ and $0.35$. Their findings are generally
consistent with our results, and they also make the important point
that this appears to be an example of deconfined quantum
criticality.


\appendix
\section{Appendix A. Ground state energy per site and the spin gaps for $J_2=0.3$.}

\begin{figure}
\includegraphics*[width=6cm, angle=0]{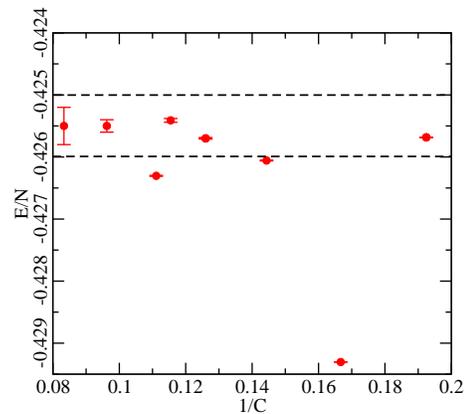}
\caption{The ground state energy per site at $J_2=0.3$ for various
cylinders with circumferences $C$. The two dashed lines are
$E/N=-0.4250$ and $-0.4260$, which are the approximate upper and
lower bounds of ground state energy that we estimate for the
infinite two-dimensional system.}\label{ener}
\end{figure}

In this section, we present the ground state energies for $J_2=0.3$
for various cylinders in Fig. \ref{ener}. The energy per site for a
given cylinder is calculated by subtracting the energies of two long
cylinders with different lengths \cite{rev}. We find that the energy
is much lower on the narrow YC4-0 cylinder ($C=6$), probably due to
the contribution to the energy from short resonating paths that wind
around the cylinder. This effect persists to a lesser extent for
YC6-0 ($C=9$). On all the other cylinders, the energy is
consistently around -0.4255(5) with wider cylinders having larger
uncertainty due to the extrapolation to large $m$. Thus we estimate
an approximate upper bound on the energy as -0.4250. This bound is
lower than other recent bounds from the variational entangled pair
states method ($E\leq -0.4210$) \cite{eps1,foota} and VMC ($E\leq
-0.4169$) \cite{vmc1}.

To explore the low-lying excited states on these cylinders, we apply
DMRG to find the singlet and triplet gap. We lock the PVB pattern on
the cylinder ends to guarantee that the excited state doesn't
include translations of the PVB pattern. For the triplet gap, we
target the ground state in the $S_z=0$ and $S_z=1$ sectors
separately. We make sure the states have the expected properties,
e.g. the $S_z=0$ state has no signs of excitations and the $S_z=1$
state has the excitation in the bulk, not at an edge. For the
singlet gap, we target the ground state in the $S_z=0$ sector first,
until this state is very accurate. Then we target one more state in
the same sector, with the sweeping restricted to exclude the end
regions. The advantage is that the singlet excitations will always
appear in the center of cylinder. We perform the calculation on long
cylinders to make sure that the gap is well converged. We use length
20 for XC cylinders and length 30 for YC cylinders.

\begin{figure}
\includegraphics[width=6cm, angle=0]{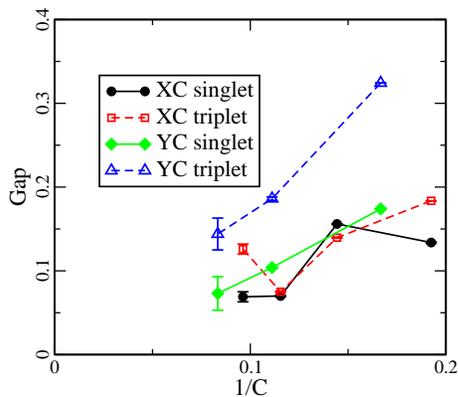}
\caption{(color online) Spin singlet (solid line with filled
symbols) and triplet (dashed line with open symbols) gaps for
various cylinders versus inverse of circumference at $J_2=0.3$.
}\label{gap}
\end{figure}

The triplet and singlet gaps for XC and YC cylinders are shown in
Fig. \ref{gap}. The triplet gap decreases as the cylinders get
wider, with the YC cylinders having higher triplet gaps than the XC
cylinders. The triplet excitations on the XC cylinders appear to
consist of two well-separated spinons. This effect is more
pronounced when the cylinder gets wider. In YC cylinders, it appears
that the spinons remain tightly bound. We don't yet understand why
the behavior is so different between the XC and YC cylinders. From
the trend of the triplet gap with circumference, particularly for
the YC cylinders, we estimate that in 2D limit the triplet gap is no
more than $0.12$. The singlet gaps also show substantial variation
between different cylinders. From the trends with circumference, we
estimate that in the 2D limit the singlet gap is no more than
$0.07$.

\section{Appendix B. Even and odd topological sectors on XC8-0 and XC10-0 cylinders.}

\begin{figure}
\includegraphics*[width=8cm, angle=0]{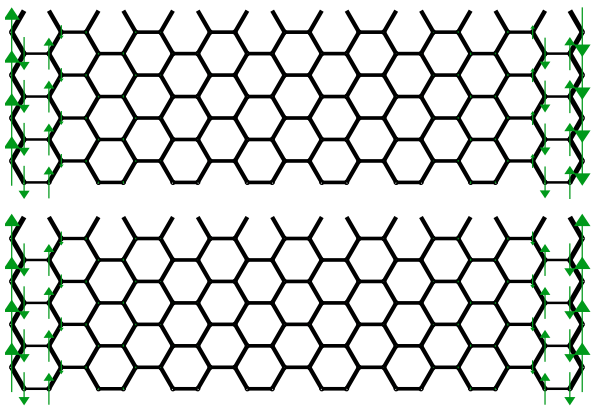}
\caption{Top panel shows the ground state of a XC8-0 cylinder for
$J_2=0.3$. It has a spinon with $S_z=1/2$ and an anti-spinon with
$S_z=-1/2$ on the edges. The bottom panel shows the excited state in
the $S_z=1$ sector with two spinons on the edges. These states have
almost exactly the same energy. }\label{xc8}
\end{figure}

In this section, we will discuss the XC8-0 and XC10-0 cylinders in
detail.  These cylinders are not compatible with PVB order, and show
interesting behavior. The frustration of the PVB order leaves two
different nearly-degenerate ground states corresponding to different
topological sectors. The sectors can be selected by adding or
removing sites from the left and right edges. For zigzag-column left
and right edges, the XC8-0 and XC10-0 ground states have two free
spinons on the cylinder edges, leading to four nearly degenerate
states (Fig. \ref{xc8}). The edge spinons can be thought of as
unpaired spins in a resonating valence bond picture.
Nearest-neighbor dimer coverings of a cylinder are in two distinct
topological classes, specified by whether a vertical cut through the
cylinder breaks an even or odd number of dimers.  We can remove one
site from each end of an XC cylinder to eliminate the end spinons.
Since there are then N-1 sites remaining on the edge of XCN
cylinder, this forces the dimer cover to be in the odd sector.

The opposite topological sector also can be produced by adding two
vacancies inside the cylinder. The effect of vacancies is to force
the the opposite topological sector between the two vacancies. The
two vacancies need to sit in special locations and not too far
apart; otherwise spinons may bind to the vacancies, thus keeping the
system in the same sector. (not shown)

\begin{figure}
\includegraphics*[width=8cm, angle=0]{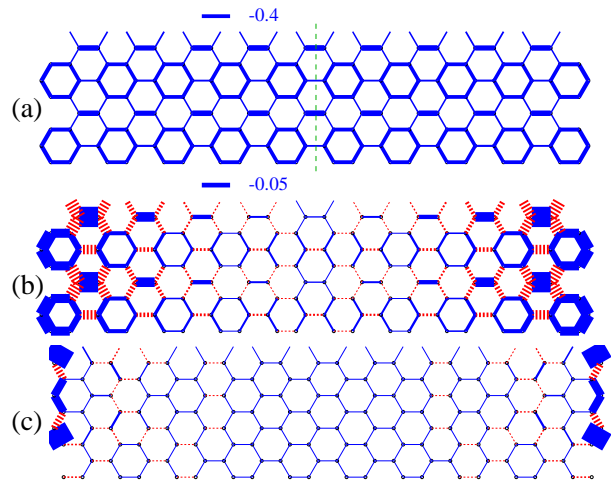}
\caption{(color online) (a) Initial state for the even topological
sector of the XC8-0 cylinder at $J_2=0.3$. The width of the lines
are proportional to $|\langle S_i\cdot S_j\rangle-e|$ with e=0. A
vertical line cuts through two singlet bonds. (b) The VBC ground
state for the even sector with e=-0.32. (c) The spin-liquid-like
ground state for the odd sector with e=-0.29 for horizontal bonds
and e=-0.34 for vertical bonds. The spin correlations are fairly
uniform in the center. On this XC8-0 cylinder the absolute ground
state is that of the odd sector.}\label{xc8eo}
\end{figure}

For the XC8-0 cylinder, the higher energy sector is the even
topological sector, which is related to a VBC with an 8 site unit
cell with a hexagon and a dimer, shown in the middle panel of Fig.
\ref{xc8eo}. To get the energy per site for the even sector, we cut
the ends and strengthen dimer bonds to favor the even sector
initially (Fig. \ref{xc8eo}a). Then we remove these strengthened
bonds to run the DMRG to get the final states (Fig. \ref{xc8eo}b).
The energy splitting for the XC8-0 cylinder is shown on the left two
panels of Fig. \ref{split}. The odd sector energy is always lower in
the intermediate phase. This state looks more uniform (Fig.
\ref{xc8eo}c) than the even sector, which shows strong valence bond
order, but it has significant anisotropy. In the intermediate phase
regime, the energy splitting increases with $J_2$, then decreases as
$J_2$ approaches the second phase transition point. We didn't
include results from higher $J_2$ values, since the energy splitting
is already small compared to the uncertainty in the extrapolation of
the energy. The fact that the energy splitting reduces with $J_2$ as
one approaches the second phase transition point means that the
phase transition in to the SVBC phase happens at essentially the
same $J_2$ value in either sector. In Fig. 3 of the main text the
odd sector was used, but we also did the same analysis in the even
sector, obtaining the same estimates of $J_2$ at the phase
transition.

\begin{figure}
\includegraphics*[width=8cm, angle=0]{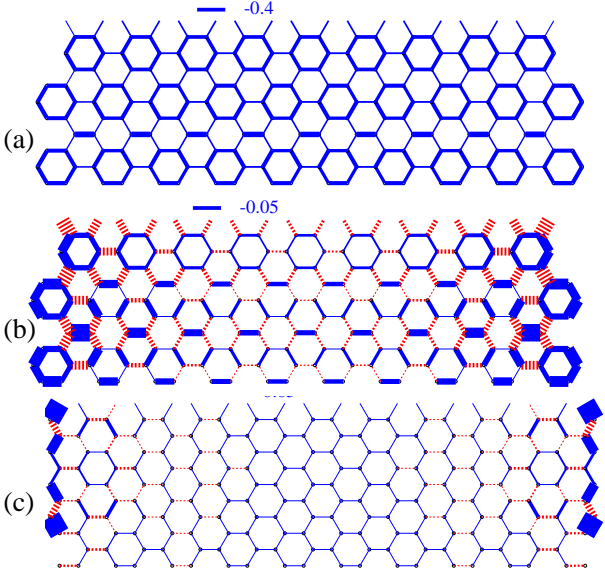}
\caption{(color online) Similar to previous figure, (a) and (b) are
the initial and the ground state for the even topological sector of
the XC10-0 cylinder with $J_2=0.3$. (c) is the odd topological
sector.  Unlike the XC8-0 cylinder, the even sector with apparent
valence bond order is actually the ground state. }\label{xc10eo}
\end{figure}

\begin{figure}
\includegraphics*[width=8cm, angle=0]{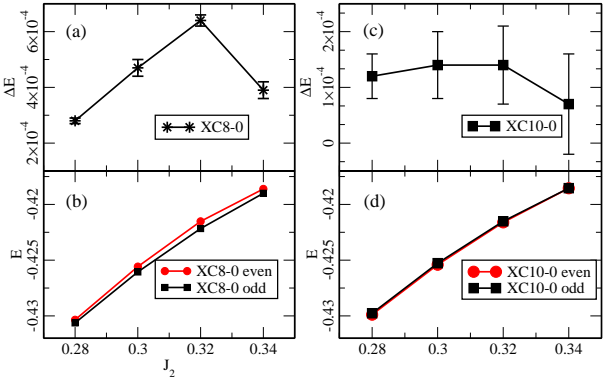}
\caption{(color online) Different topological sector energy
splittings for the XC8-0 and XC10-0 cylinders versus $J_2$ with
error bars. (a) and (c) are the energy splittings for the XC8-0 and
XC10-0 cylinders respectively. (b) and (d) are the energy per site
for both even and odd sectors on XC8-0 and XC10-0 cylinders. It's
clear that on the XC8-0 cylinder the odd sector energy is lower. On
the XC10-0 cylinder, the energy splitting is quite small, and
becomes consistent with zero at $J_2=0.34$.}\label{split}
\end{figure}

The even sectors of XC10-0 cylinder are shown on Fig. \ref{xc10eo}.
The even sector has a 20 sites VBC unit cell. The final state has
stronger valence bond order than the XC8-0 cylinder. As opposed to
XC8-0 cylinder, we show in the right two panels of Fig.\ref{split}
that the even sector is lower in energy. We could measure the spin
correlation length from placing staggered magnetic fields on the
edge and measuring how $S_z$ decays. The odd sector has a much
longer spin correlation length with $\xi_s\sim12$, as compared with
roughly 3 in even sector. A longer spin correlation length usually
indicate smaller spin triplet gap. We find that spin triplet gap is
roughly $10^{-3}$ for the odd sector compared with 0.075 for the
even sector. Thus these results indicate that on XC10-0 cylinder the
ground state is a valence bond crystal with short range spin
correlation and a nonzero spin triplet gap.

\begin{figure}
\includegraphics*[width=8cm,angle=0]{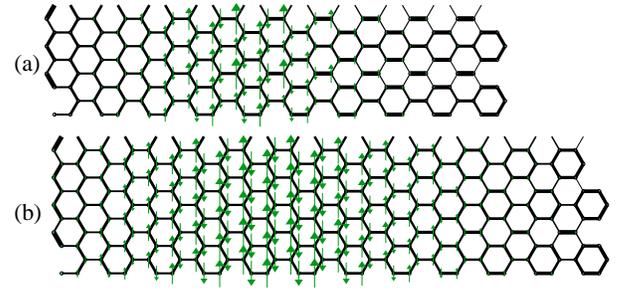}
\caption{(a) is XC8-0 cylinder with delta bond applied from 0.02
(left edge) to -0.05(right edge), with positive delta strengthen the
vertical bonds and negative delta value strengthen the horizontal
bonds. Thus we cut the cylinder edge to favor odd sector for
positive delta and even sector for negative delta. Spinons locate
between even and odd sectors. (b) is XC10-0 cylinder with delta from
0.01 to -0.01. }\label{oe}
\end{figure}

\begin{figure}
\includegraphics*[width=6cm,angle=0]{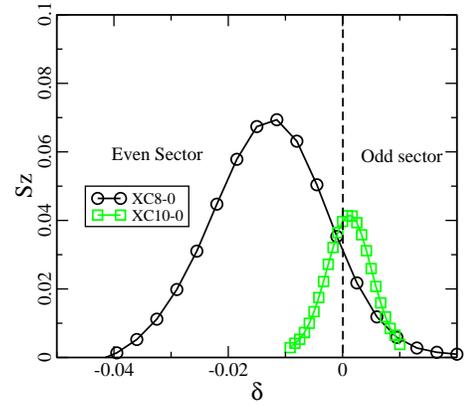}
\caption{Delta value versus sum of $S_z$ along the cylinder zigzag
direction for XC8-0 and XC10-0 cylinder. The data are extracted from
previous figure. The left side of spinon ($\delta<0$) is even
sector, with odd sector on the spinon right side. We could see that
for ground state with $\delta=0$, XC8-0 is in odd sector and XC10-0
is in even sector. }\label{spinonoe}
\end{figure}

To further check that the even sector has lower energy, we
strengthen and weaken some bonds to favor both sectors on one
cylinder; one on the left, the other on the right, see Fig.
\ref{oe}. A domain wall with spinon will reside between these
different sectors. Depending on whether the spinon is more pushed
into the even or the odd sector at $\delta=0$, we can determine
which sector is the ground state. The results are consistent on both
cylinders, shown in Fig. \ref{spinonoe}.

\section{Appendix C. Entropy analysis for intermediate phase with $J_2=0.3$.}

In this section, we will discuss the von Neumann entanglement
entropy (EE) in detail for $J_2=0.3$ for all the XC and YC
cylinders. For a cylinder with a non-shifted connection, eg. the
XCN-0 and YCN-0 cylinders, we calculate the EE for splitting the
lattice into subsystems A and B. The partition is a vertical line
cut through the cylinder, that always breaks the same number of
horizontal or diagonal bonds, i.e. breaking N/2 bonds for XCN-0
cylinder and N bonds for YCN-0 cylinder. We didn't consider the EE
for cylinders with shifted connections.

In a 2D disordered gapped system, the EE obeys the area law \cite{et0}
with
\begin{equation}
S=aL+o(1/L)
\end{equation}
where $a$ is a non-universal constant, L is the length of the
boundary between regions A and B. The second term disappears as
$L\rightarrow \infty$. For a gapped topologically-ordered SL phase
without broken symmetries, a extra term is included
\begin{equation}
S=aL+\gamma+o(1/L)
\end{equation}
with $\gamma$ is the topological EE\cite{et4, et5}. For a minimum
entangled state with $Z_2$ topological order, $\gamma=-\ln(2)$ at
$T=0$. Thus the EE is a constant independent of length for a fixed
width cylinder, since the boundary length is always the same. For a
2D critical gapless system, the EE scales as:
\begin{equation}
S=aL+b\ln{L}+c(L)\ln[\sin(\frac{\pi x}{L})]+o(1/L)
\end{equation}
for a $L\times L$ torus\cite{et1, et2,et3}. The second term comes
from gapless Goldstone mode. $x$ is subsystem size and the third
term is the function depends on ``chord length''. Thus the EE
depends on system width L and subsystem size $x$ for a 2D gapless
system, but independent of subsystem size for a gapped system.

\begin{figure}
\includegraphics*[width=8cm, angle=0]{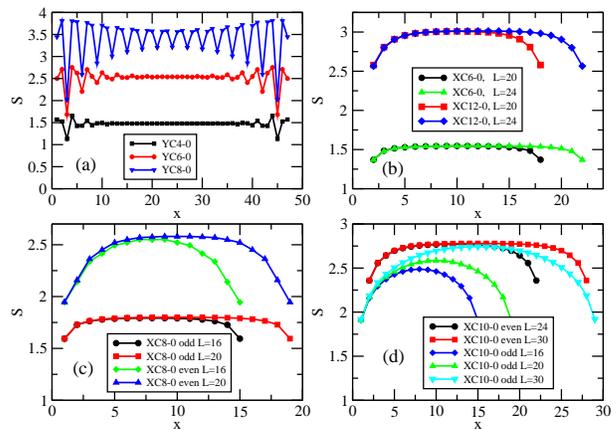}
\caption{(color online) The EE versus subsystem size $x$ for (a) YC
cylinders, (b) XC6-0 and XC12-0 cylinders, (c) even and odd sector
in XC8-0 cylinders, (d) even and odd sector in XC10-0 cylinders.
Note on YC8-0 cylinder, the EE is measured for a state with
$m=5000$. This state is not well converged and it has larger PVB
order as seen in the EE oscillation pattern. }\label{xycs}
\end{figure}

In Fig. \ref{xycs}(a), we show the EE versus subsystem size $x$ for
YCN-0 cylinders. Since YC cylinders have strong plaquettes appearing
on the cylinder edges, we would expect that the EE should oscillate
with a PVB pattern, with two high values and one low value
corresponding to cuts at strong and weak bonds of plaquettes. With
the decay of PVB order to the center, we also observe the decay of
the EE oscillation pattern and measure the EE when its value
saturates to a constant value.  We notice that YC6-0 has a longer
PVB correlation length than YC4-0, since the EE oscillates with more
periods near the YC6-0 cylinder edges. On the YC8-0 cylinder, the
state in which we measure the EE is not well converged, and thus it
has stronger PVB order. The EE shows an apparent PVB order
oscillation pattern.

For the XC cylinders in Fig. \ref{xycs}(b-d), the entropy is
saturated in the center, independent of subsystem size for the
ground state. For the XC8-0 cylinder, the odd sector EE is
independent of the sub-system size $x$ and has the same value for
cylinders of different lengths. The EE has much stronger dependence
on the sub-system size for the even sector. On the contrary for the
XC10-0 cylinder, the EE saturates in the center of the cylinder for
the even sector. All these results again show that the GS for all
the cylinders with $J_2=0.3$ are gapped with short spin correlation
lengths.

\begin{figure}
\includegraphics[width=6cm, angle=0]{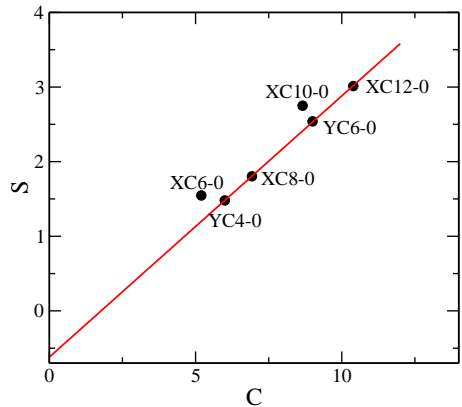}
\caption{Saturated EE versus cylinder circumference C for the XC and
YC cylinders. }\label{e-w}
\end{figure}

In Fig. \ref{e-w}, we present the EE versus cylinder width to
extrapolate the topological EE $\gamma$. It is surprising that not
all the data points collapse onto a line. The EE from XC6-0 and
XC10-0 deviates from a linear extrapolation. It is possible that
XC6-0 cylinder is too narrow, so that $J_2=0.3$ is too close to the
second phase transition point. The XC10-0 cylinder is a valence bond
crystal with 20 sites unit cell in the even sector. We find that
$\gamma=-0.619$, which is close to gapped $Z_2$ SL with
$\gamma=\ln2=-0.693$.

In a recent paper by Hong-Chen Jiang et al.\cite{tee1}, the authors
extrapolate the EE to get the TEE with $\gamma\sim-\ln2$ and 
identify $Z_2$ SL phase on many models. As those authors realize,
for the above method to be accurate when identifying topological
phases, all the correlation lengths (spin-spin, dimer-dimer, etc.)
have to be short compared with the cylinder widths, so that DMRG
results will always lead to a minimum entangled state (MES) with
zero vison flux through the cylinder. For the model in this paper,
even though the spin-spin correlation length is short, PVB
correlation lengths get longer as the cylinder gets wider. Thus the
extrapolation method for the TEE would lead to strong finite size
effects. Therefore the non-zero $\gamma$ we find in this paper
doesn't indicate that the GS for the Heisenberg $J_1-J_2$ model on
the honeycomb lattice is a $Z_2$ SL in the 2D limit.


\begin{thebibliography}{36}
\bibitem[{foo({\natexlab{a}})}]{foot}
\bibinfo{journal}{\text{Note} that the six spin interactions may be more
  important than the $J_2$ terms in properly characterizing the charge
  fluctuations in the Hubbard model \cite{eff1, eff2}}.
\end{thebibliography}
\begin{thebibliography}
\bibitem[{foo()}]{foota}
\bibinfo{note}{We use table I of Ref. \cite{eps} at $J_2=0.3$. We extract the
  data from L=9, 16, 18 and extrapolate versus $L^{-3/2}$ to get the energy in
  the 2D limit as $L\rightarrow\infty$.}
\end{thebibliography}
\end{document}